\documentclass[useAMS]{mn2e}
\usepackage{rotating}
\usepackage{mncite}

\def\approxgt{\ifmmode \rlap{$>$}{}_{{}_{{}_{\textstyle\sim}}} \else%
$\rlap{$>$}{}_{{}_{{}_{\textstyle\sim}}}$\fi} 
\def\approxlt{\ifmmode \rlap{$<$}{}_{{}_{{}_{\textstyle\sim}}} \else%
$\rlap{$<$}{}_{{}_{{}_{\textstyle\sim}}}$\fi}

\LARGE \normalsize \title{MS~1603.6+2600: an Accretion Disc Corona
  source?}

\author[P.G. Jonker et al.]  {P.G. Jonker$^1$\thanks{email
    :peterj@ast.cam.ac.uk}, M. van der Klis$^2$, C. Kouveliotou$^3$,
  M.  M\'endez$^4$, W.H.G. Lewin$^5$, \newauthor
  T. Belloni$^6$\\
  $^1$Institute of Astronomy, Madingley Road, CB3 0HA, Cambridge\\
  $^2$Astronomical Institute ``Anton Pannekoek'',
  University of Amsterdam, Kruislaan 403, 1098 SJ Amsterdam\\
  $^3$SD 50, Space Science Research Center, National Space Science and Technology Center, 320, Sparkman Drive, Huntsville, AL 35805, \\Nasa's Marshall Space Flight Center\\
  $^4$SRON, National Institute for Space Research, Sorbonnelaan 2,
  3584~CA
  Utrecht, The Netherlands\\
  $^5$Department of Physics and Center for Space Research, Massachusetts Institute of Technology, Cambridge, MA 02138\\
  $^6$INAF, Osservatorio Astronomico di Brera, Via E. Bianchi 46,
  23807 Merate, Italy}

\begin{document}

\maketitle

\begin{abstract}
\noindent 
We have observed the eclipsing low--mass X--ray binary MS~1603.6+2600
with {\it Chandra} for 7 ksec. The X--ray spectrum is well fit with a
single absorbed powerlaw with an index of $\sim$2. We find a clear
sinusoidal modulation in the X--ray lightcurve with a period of
1.7$\pm0.2$ hours, consistent with the period of 1.85 hours found
before. However, no (partial) eclipses were found. We argue that if
the X--ray flare observed in earlier X--ray observations was a type I
X--ray burst then the source can only be an Accretion Disc Corona
source at a distance between $\sim$11--24 kpc (implying a height above
the Galactic disk of $\sim$8--17 kpc). It has also been proposed in
the literature that MS~1603.76+2600 is a dipper at $\sim$75 kpc. We
argue that in this dipper scenario the observed optical properties of
MS~1603.6+2600 are difficult to reconcile with the optical properties
one would expect on the basis of comparisons with other high
inclination low--mass X--ray binaries, unless the X--ray flare was not
a type I X--ray burst. In that case the source can be a nearby soft
X--ray transient accreting at a quiescent rate as was proposed by
Hakala et al.~(1998) or a high inclination source at $\sim$15--20 kpc.

\end{abstract}


\section{Introduction}
\label{intro}
Low--mass X--ray binaries (LMXBs) are binary systems in which a
$\approxlt$1\,$M_{\odot}$ star transfers matter to a neutron star or a
black hole. A large fraction of these LMXBs are found in the Galactic
Bulge (see \pcite{vawh1995}). MS~1603.6+2600 was discovered with the
{\it Einstein} satellite as a faint source at a flux level of
$\sim1\times10^{-12}{\rm erg\,cm^{-2}\,s^{-1}}$ (0.3--3.5 keV;
\pcite{1990ApJS...72..567G}).  \scite{1990ApJ...365..686M} found an
optical counterpart that shows partial eclipses with a period of 111
minutes; they also reported that the depth of the optical eclipse is
anti--correlated with the optical luminosity. No radio emission was
detected with an upper limit of 0.3 mJy (\pcite{1990ApJ...365..686M}).
{\it ROSAT} and {\it ASCA} detected this source at flux levels of
$\sim1\times10^{-12}{\rm erg\,cm^{-2}\,s^{-1}}$ (0.1--10 keV) and
$\sim4\times10^{-12}{\rm erg\,cm^{-2}\,s^{-1}}$(0.7--10 keV),
respectively (\pcite{1990ApJ...365..686M}; we determined the flux from
the spectral properties and count rate measured with {\it ROSAT};
\pcite{1998A&A...333..540H}).
  
It was immediately realized that this system is either a Cataclysmic
Variable (the compact object is a white dwarf) or an LMXB (the compact
object is a neutron star or a black hole) and that if this system is
an LMXB its distance may be large (\pcite{1990ApJ...365..686M}).
However, \scite{1998A&A...333..540H} proposed that the system could be
a nearby soft X--ray transient in quiescence accreting at a low rate.
They also showed that the system properties do not fit any of the
Cataclysmic Variable categories.  \scite{1993A&A...277..483E} showed
that there are three possible evolutionary scenarios for this system
and they argue that the system contains a neutron star rather than a
white dwarf compact object.  \scite{2001ApJ...561..938M} detected an
X--ray flare, which could be a type~I X--ray burst.  If the flare was
indeed a type~I X--ray burst it establishes the nature of the compact
object as a neutron star.

In this Paper we report on a {\it Chandra} observation of
MS~1603.6+2600.

\section{Observations and analysis}
\label{analysis}

The {\it Chandra} satellite (\pcite{1988SSRv...47...47W}) observed
MS~1603.6+2600 with the Advanced CCD Imaging Spectrometer (ACIS) on
March 11, 2002 for nearly 7 ksec starting at 08:19 (TT; MJD
52344.346). The data were processed by the {\it Chandra} X--ray
Center; events with ASCA grades of 1, 5, 7, cosmic rays, hot pixels,
and events close to CCD node boundaries were rejected. We used the
standard {\it CIAO} software to reduce the data (version 2.3 and CALDB
version 2.21). A very weak streak caused by the arrival of photons
during the CCD readout period (40 $\mu$seconds) was present. We
removed this streak before processing the data further using the {\sl
  ciao} tool {\sc acisreadcorr} taking into account that we had
windowed the CCD to 1/4 to reduce effects of pile--up.

We detect only one source. After applying the {\sl ciao} web--based
tool {\sc fix offsets} to correct for known aspect offsets, we derive
the following coordinates for the source: R.A.=16h05m45.88s,
Decl.=+25$^\circ$51'45.1" (typical error 0.6", equinox 2000.0). The
USNO B1.0 coordinates (\pcite{2003AJ....125..984M}) of the optical
counterpart of MS~1603.6+2600 [UW Coronae Borealis; USNO B1.0 Id.
1158--0232558, R.A.=16h05m45.868(3)s Decl.=+25$^\circ$51'45.560(3)"]
are fully consistent with this.

\begin{figure*}
  \includegraphics[width=10cm]{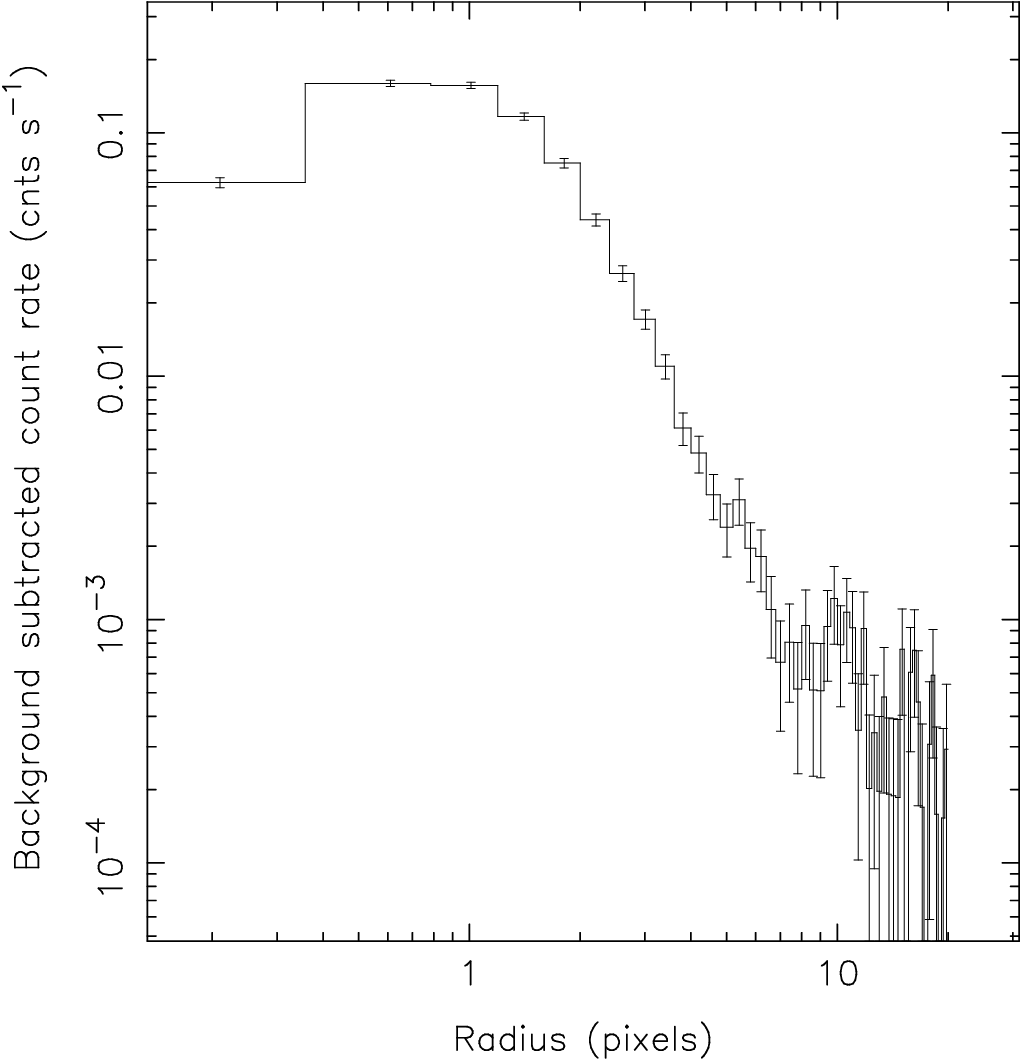}
\caption{Radial profile of the source. The effect 
  the pile--up has on the radial distribution of the point spread
  function is clearly visible as a reduction of the count rate within
  the inner pixel.}

\label{radprf}
\end{figure*}

The source is detected at a count rate of 0.72$\pm$0.01 counts per
second; this could yield a pile--up fraction of approximately 25--30
per cent for our frame time of 0.84~s.  In order to investigate this
further, we study the radial profile of the point spread function of
the source (see Figure~\ref{radprf}).  Comparing the profile
(Figure~\ref{radprf}) with the theoretical profiles shown in figure
6.24 of the {\it Chandra} Proposers' Observatory Guide v.5
\footnote{available at http://asc.harvard.edu/udocs/docs/docs.html} we
note that the number of counts per pixel is lower in the inner 1--2
pixels than it would have been in the absence of pile--up. 

The spectra are extracted with 20 counts per bin. We only include
energies above 0.3 and below 8 keV in our spectral analysis since the
ACIS timed exposure mode spectral response is not well calibrated
below 0.3 keV and above 8 keV. We fit the spectra using {\sc XSPEC}
(\pcite{1996adass...5...17A}) version 11.2.0ao, including an extra
multiplicative model component in the fit--function, the {\sc ACISABS}
model \footnote{see
  http://asc.harvard.edu/cal/Acis/Cal\_prods/qeDeg/}, to correct for
additional absorption due to contamination by the optical blocking
filters in all our spectral fits. Since this model is only accurate to
approximately 10 per cent we furthermore include a 10 per cent
systematic uncertainty to the channels below 1 keV (channels 1--69).
To correctly model the effect of pile--up on our observed spectrum, we
fitted the spectrum using the {\sc pile--up} model of
\scite{2001ApJ...562..575D}.

\begin{figure*}
  \includegraphics[width=10cm,angle=-90]{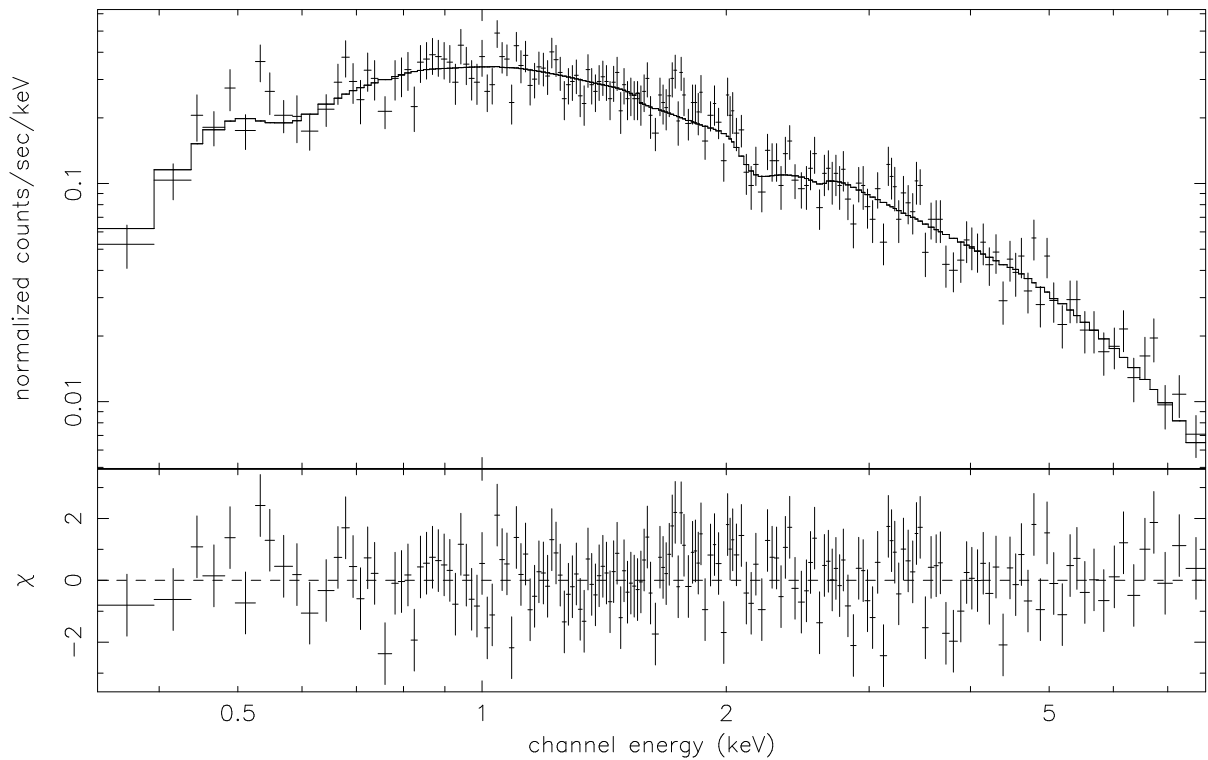}
\caption{Upper panel: {\it Chandra} ACIS--S X--ray spectrum (0.3--8 keV) 
  of the source MS~1603.6+2600. The best--fit power--law model
  modified by the combined effects of pile--up (Davis 2001) and
  interstellar absorption and absorption due to contamination of the
  optical blocking filters of the ACIS instrument derived by using
  XSPEC is overplotted. Lower panel: data minus model residuals. Since
  the residuals are divided by the error the Y--axis is dimensionless
  (the size of the error bars is unity).}
\label{spec}
\end{figure*}

The spectrum is not well fit by a single absorbed blackbody component
[reduced $\chi^2=2.9$ for 173 degrees of freedom (d.o.f.)].  A single
absorbed power law yields a good fit, with reduced $\chi^2=1.0$ for
173 d.o.f. (see Figure~\ref{spec}). The power law index is 2.0$\pm$0.1
for an interstellar absorption, $N_H$, of
$(1.5\pm0.2)\times10^{21}{\rm cm}^{-2}$. This $N_H$ is inconsistent
with the value derived by \scite{1990ARA&A..28..215D}
($\sim4\times10^{20}{\rm cm}^{-2}$).  However, using ASCA data
\scite{2001ApJ...561..938M} find that the absorbing column to this
source varies as a function of the binary orbital phase; it can be as
high as 4.6$\times10^{21}{\rm cm}^{-2}$.  The unabsorbed flux (0.1--10
keV) is 1.2$\times10^{-11}$ erg\,cm$^{-2}$\,s$^{-1}$. We present the
parameters for this model in Table~\ref{specpars}. To test the
consistency of the XSPEC pile--up model implementation we also fitted
the spectrum with the {\sc ISIS} package version 1.1.3.
(\pcite{2000adass...9..591H}) and their implementation of the pile--up
model of \scite{2001ApJ...562..575D}. We find that the results are
consistent within the 90 per cent uncertainties. We note that in the
ISIS fits we corrected the auxiliary response file using the CIAO tool
{\sl corrarf} to take the degredation of the ACIS quantum efficiency
due to the contamination of the optical blocking filters into account.
We present the ISIS fit--results in the second line of
Table~\ref{specpars}.

\begin{table*}
\caption{Best fit parameters of the spectra of MS~1603.6+2600. All quoted 
errors are at the 90\% confidence level. In 
deriving the value for the interstellar absorption, the local absorption 
due to the {\it Chandra} optical blocking filters was accounted for. The first 
line describes the best--fit parameters obtained when using {\sc XSPEC} 
whereas on  the second line the best--fit parameters are given when using 
{\sc ISIS}.}
\label{specpars}
\begin{center}

\begin{tabular}{ccccc}
\hline
N$_H$  & PL$^a$ & & Pile--up parameter & Reduced \\
($\times10^{21}$ cm$^{-2}$) &  Index  &  10$^{-3}$ photons keV$^{-1}$ cm$^{-2}$ s$^{-1, b}$ & $\alpha$$^c$& $\chi^2/$d.o.f.\\
\hline
\hline
1.5$\pm$0.3 & 2.05$\pm$0.15 & 2.2$^{+3}_{-0.4}$ & 0.76$\pm$0.25 & 1.0/173 \\
1.2$\pm$0.3 & 1.9$\pm$0.2 & 2$\pm$3 &  0.77$\pm$0.41 & 1.1/184 \\

\end{tabular}
\end{center}

{\footnotesize $^a$ PL = power law}\newline
{\footnotesize $^b$ Power law normalisation at 1 keV.}\newline
{\footnotesize $^c$ Parameter has a hard upper limit of 1.}\newline
\end{table*}

\begin{figure*}
  \includegraphics[width=10cm]{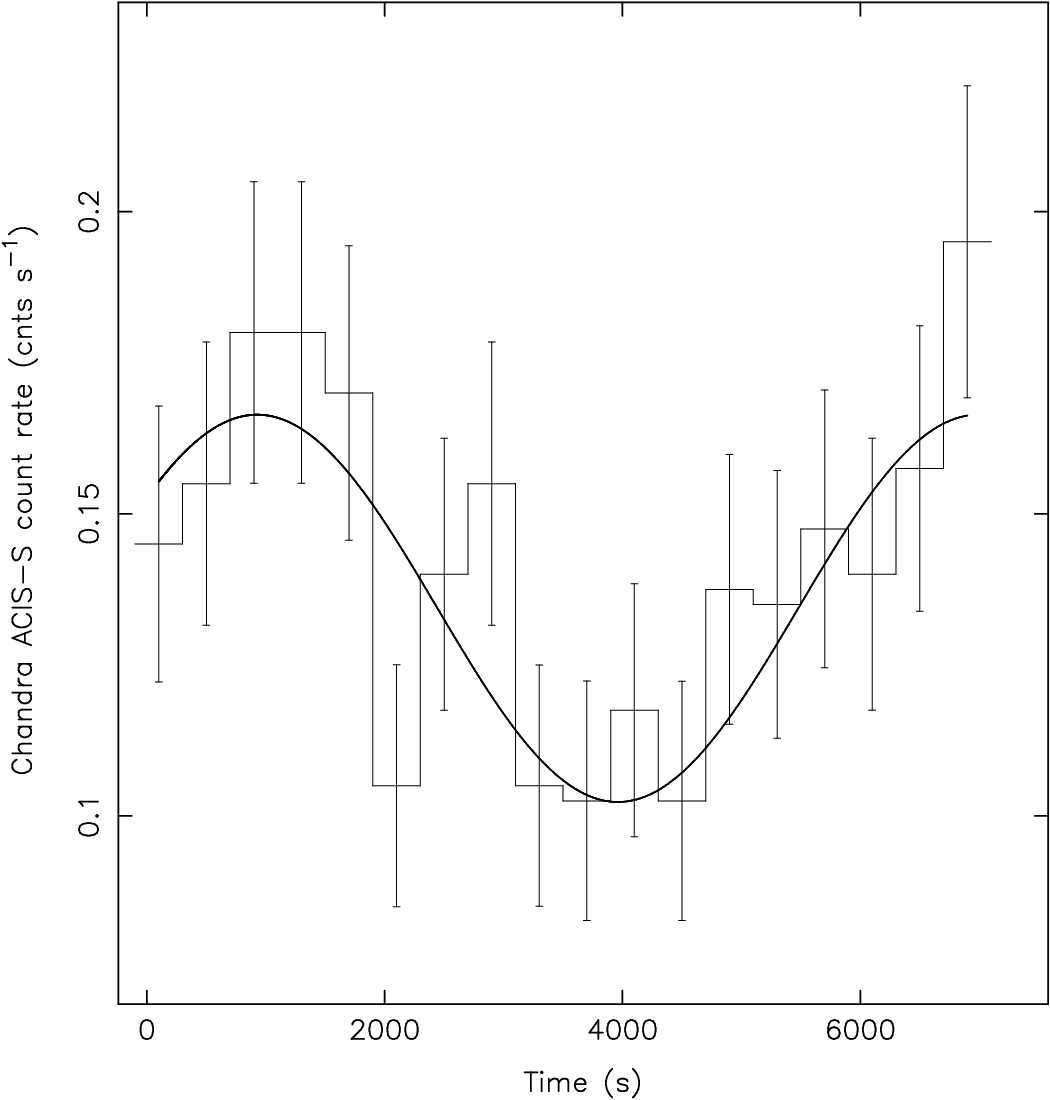}
\caption{Lightcurve of MS~1603.6+2600 in the 0.3--10 keV energy 
  band, excluding data from a circular region with a 2 pixel radius
  centred on the position of MS~1603.6+2600.  Each bin is an average
  of 400 seconds of data. Clearly visible is the sinusoidal variation
  at the orbital period of $\sim$1.85 hours ($\sim$6600 seconds). The
  solid line is the best--fit sinusoid to the data, with a period of
  1.7$\pm$0.2 hours. Time zero is MJD 52344.346. Note that the
  background is not subtracted.}
\label{lcs}
\end{figure*}

We create lightcurves with a time resolution of 10 and 200 seconds
using photons with energies 0.3--10 keV excluding a circular area of a
2 pixel radius centred on the source position to mitigate the effect
the pile-up has on the lightcurve. The lightcurve with a resolution of
10~s was searched for the presence of type I X--ray bursts, but none
was found. In Figure~\ref{lcs} we plot the lightcurve in the 0.3--10
keV energy band. We could not phase connect our observation with
previous observations since due to the uncertainties in the ephemeris,
and the long period between our observations and the zero point of the
ephemeris (\pcite{1990ApJ...365..686M}) the cycle count is lost. A
clear sinusoidal variation is present; a fit of a sinusoid gives a
period of 1.7$\pm0.2$ hours, consistent with the period of 1.85 hours
found before (\pcite{1990ApJ...365..686M};
\pcite{2001ApJ...561..938M}). We produced lightcurves in different
energy bands to search for variations in the amplitude of this
modulation (0.3--1.5 keV, 1.5--10 keV). We fixed the period and the
phase of the sinusoid to the values obtained from the fit to all
energy bands combined. The amplitude in the 0.3--1.5 keV band was
(2.0$\pm0.5$)$\times10^{-2}$ counts s$^{-1}$ whereas that in the
1.5--10 keV band was ($1.3\pm0.5$)$\times10^{-2}$ counts s$^{-1}$
(errors are 1$\sigma$ single parameter errors). We note that the
25--30 per cent pile--up could in principal have affected the
amplitude of the sinusoidal modulation as a function of energy.

\section{Discussion}

We observed MS~1603.6+2600 with the ACIS--S instrument onboard the
{\it Chandra} satellite for nearly 7 ksec. Only one source was
detected and its position is consistent with the USNO B1.0 position of
the optical counterpart. The source was piled--up but still useful
spectroscopic parameters could be derived by using the pile--up model
of \scite{2001ApJ...562..575D}. The best--fit spectrum is that of an
absorbed power law with photon index $\sim$2 and an interstellar
absorption consistent with the values found by
\scite{2001ApJ...561..938M}. The power law index of 2 is typical for
a low--mass X--ray binary (LMXB). \scite{1990ApJ...365..686M} reported
that MS~1603.6+2600 has a hard spectrum using {\it Einstein}
observations.

The lightcurve shows a sinusoidal modulation with a period consistent
with that derived previously from optical photometric and X--ray
observations (\pcite{1990ApJ...365..686M};
\pcite{2001ApJ...561..938M}). Similar to the 1991 {\it ROSAT PSPC}
observations (\pcite{1998A&A...333..540H}), no partial X--ray eclipse
is present during our 2002 {\it Chandra} observation.  However,
eclipses have been observed during a 1997 {\it ASCA} observation.
\scite{2001ApJ...561..938M} found a clear dependence of the lightcurve
profile on the photon energy. They observed an eclipse at X--rays
($<$2 keV) whereas at energies $>$2 keV the eclipse is much less
pronounced; we find no evidence for a change in the amplitude of the
sinusoidal lightcurve profile as a function of energy.

Perhaps the fact that we do not find evidence for such behaviour ties
in with the behaviour observed at optical wavelengths.  If the overall
source luminosity is higher, the eclipses in optical are less
pronounced; at the highest optical luminosities they disappear
completely (\pcite{1990ApJ...365..686M}).  Indeed, the 0.1--10 keV
unabsorbed flux at the epoch of the observations of
\scite{2001ApJ...561..938M} ($\sim$5--8$\times10^{-12}\,{\rm
  erg\,cm^{-2}\,s^{-1}}$) is a factor of 1.5 lower than the unabsorbed
0.1--10 keV flux we measured in our {\it Chandra} observations
(1.2$\times10^{-11}$ erg\,cm$^{-2}$\,s$^{-1}$).

As mentioned in the introduction it has been considered that
MS~1603.6+2600 is a Cataclysmic Variable (CV) and not an LMXB.
However, as discussed in \scite{1998A&A...333..540H} and
\scite{2001ApJ...561..938M} the optical and X--ray properties of
MS~1603.6+2600 are difficult to reconcile with a CV scenario. So, if
MS~1603.6+2600 is a CV it must be an unusual one
(\pcite{2001ApJ...561..938M}). There are three possible LMXB scenarios
for MS~1603.6+2600 presented in the literature; below we will discuss
all three of them.

Interpreting the observed X--ray flare as a type~I X--ray burst,
\scite{2001ApJ...561..938M} note that the source distance must be
large ($\sim$75 kpc), consistent with earlier estimates (30--80 kpc;
\pcite{1990ApJ...365..686M}). They further argue that based on the
variability of the lightcurve in optical and X--rays the source
differs from an ADC source like 2S~1822--371, and hence it is a dipper
source like 4U~1916--053.  However, variability in ADC sources is not
new. For instance, the soft X--ray transient 4U~2129+47 is an ADC
source in outburst while in quiescence full eclipses are observed
(\pcite{1994ApJ...435..407G}; \pcite{2002ApJ...573..778N}).
Furthermore, if MS~1603.6+2600 is a dipper at 75 kpc the absolute
visual magnitude, M$_V$, of the source would be $\sim$0, much larger
than that of the high inclination source in a 50 minute orbit
4U~1916--053 (M$_V=5.3$) and even larger than that of the eclipsing
source in a 3.9 hour orbital period EXO~0748--676 (M$_V=1.4$;
\pcite{1995xrb..book...58V}). The error on the estimates of these
absolute magnitudes will be smaller than $\sim1$ magnitude since the
distance estimates for these systems were obtained using radius
expansion bursts (4U~1916--053, \pcite{1988MNRAS.232..647S};
EXO~0748--676, \pcite{gohapa1986}).  Such estimates are typically
accurate to approximately 20 per cent (\pcite{2003A&A...399..663K}).
Furthermore, the reddening towards these sources is low and well
established (4U~1916--053, \pcite{1988MNRAS.232..647S}; EXO~0748--676,
\pcite{1990A&A...227..105S}). Since the absolute optical magnitude
scales with the binary orbital period and the X--ray luminosity
(\pcite{1995xrb..book...58V}), it is unlikely that the absolute
optical magnitude of the 1.85 hour system MS~1603.6+2600 is larger
than that of EXO~0748--676. On the basis of this, we conclude that it
is unlikely that MS~1603.6+2600 is a dipper at a distance as large as
75 kpc. The X--ray and optical properties of MS~1603.6+2600 are
consistent with a dipper scenario in which the distance to the source
is $\sim$15--20 kpc if either the observed X--ray flare was not a
type~I X--ray burst or the luminosity of the burst was unusually low.

\scite{1998A&A...333..540H} favour a less distant (d$\sim$0.25--2.7
kpc) source assuming MS~1603.6+2600 is a soft X--ray transient
accreting at a low rate in quiescence. Even though the X--ray spectrum
of several neutron star soft X--ray transients in quiescence is well
fit by a black body, often a power--law component is present in the
spectrum as well. The quiescent X--ray spectrum of SAX~J1808.4--3658
was well fit by a single power law (\pcite{2002ApJ...575L..15C})
similar to the X--ray spectrum of MS~1603.6+2600 albeit somewhat
harder (the power--law index for SAX~J1808.4--3658 was 1.5). The
quiescent spectrum of black hole candidate SXTs in quiescence is well
fit by a power law with an index of $\sim$2
(\pcite{2002ApJ...570..277K}).

Several systems containing a neutron star compact object accreting at
a low rate have been identified (\pcite{2002A&A...392..931C}). An
outburst has never been observed from these systems. If MS~1603.6+2600
is indeed at a distance of approximately 1 kpc then the distance
modulus is $\sim$10.  Given the observed V band magnitude of
MS~1603.6+2600 (R=19.4, V$\sim$19.7; \pcite{1990ApJ...365..686M})
M$_V$ is $\sim$10. Given the fact that many disc lines and even the
Bowen blend are observed in the spectrum (\pcite{1990ApJ...365..686M})
the disc absolute magnitude must be close to this whereas that of the
companion star must be less than 10; constraining the spectral type of
the companion star to late type M dwarfs. Such a star indeed fits the
Roche lobe assuming a neutron star mass of 1.4 M$_\odot$. The
luminosity of MS~1603.6+2600 is close to 10$^{33}$ erg s$^{-1}$ if the
distance is 1 kpc. This luminosity is typical for a neutron star LMXB
in quiescence. Using the same line of reasoning one can show that a
late type M star also nearly fits the Roche lobe of a nearby black
hole candidate (with an assumed black hole mass of 6 M$_\odot$) SXT in
quiescence.

However, strong He~II and He~I lines are present in the spectrum of
MS~1603.6+2600.  Such He features are present in the spectra of
actively accreting low--mass X--ray binaries
(\pcite{1995xrb..book...58V}) but not in the spectrum of the quiescent
soft X--ray transient Cen~X--4 (\pcite{1987A&A...182...47V};
\pcite{1989A&A...210..114C}).  In this ``quiescent soft X--ray
transient'' scenario the X--ray flare observed by
\scite{2001ApJ...561..938M} cannot have been a type~I X--ray burst for
its luminosity was much too low for a distance of $\sim$1 kpc.  X--ray
flaring in quiescence is not unfeasible since optical flares have been
found in both neutron star and black hole soft X--ray transients in
quiescence (\pcite{2002MNRAS.330.1009H}; \pcite{2003ApJ...582..369Z}).

In ADC sources only a fraction of the true source luminosity is
observed due to scattering in the ADC.  In case of 4U~2129+47 a type I
X--ray burst was observed for which the luminosity was a factor of
$\sim$500 lower than the Eddington luminosity
(\pcite{1987ApJ...313L..59G}).  Similarly, \scite{1982ApJ...257..318W}
argue that the intrinsic source luminosity of ADC sources should be
$>10^{37}\,{\rm erg\,cm^{-2}\,s^{-1}}$; this is higher than the
luminosities one would derive for ADC sources if one assumed that the
sources are viewed directly.  Finally, \scite{2001ApJ...553L..43J}
presented evidence based on the observed spin--up of the pulsar in the
ADC source 2S~1822--371 showing that the intrinsic source luminosity
is a factor of $\sim$100 higher than the observed luminosity. If we
assume that we only observe between 1/500 and 1/100 of the intrinsic
source luminosity of MS~1603.6+2600 the distance would be $\sim$11--24
kpc assuming that the intrisic burst flux (i.e. 100--500 times the
observed flux) of the burst in MS~1603.6+2600 observed by {\it ASCA}
corresponds to the Eddington luminosity (we take for the Eddington
luminosity $\sim2\times10^{38}\,{\rm erg\,s^{-1}}$; the observed burst
peak flux was $\sim3\times10^{-11}\,{\rm erg\,cm^{-2}\,s^{-1}}$;
\pcite{2001ApJ...561..938M}). Such a distance would imply an M$_V$
intermediate between that of the high inclination 50 minute binary
4U~1916--053 and that of the 3.9 hour binary EXO~0748--676, as
expected. For a distance of $\sim$11--24 kpc the system would be
$\sim$8--17 kpc above the Galactic plane. As mentioned before
\scite{2001ApJ...561..938M} dismiss the identification of
MS~1603.6+2600 as an ADC source on the ground that MS~1603.6+2600 is
different from 2S~1822--371 in many facets.  However, it is unclear if
this is enough to dismiss MS~1603.6+2600 as an ADC source since the
group of ADC sources is highly disparate. For instance, 2S~1822--371
harbors a high magnetic field neutron star
(\pcite{2001ApJ...553L..43J}), 4U~2129+47 is a transient system
(\pcite{1994ApJ...435..407G}), and 2S~0921--630 has a sub--giant
companion star (P$_{orb}=9.02$ days; \pcite{1983MNRAS.205..403B}).  

We conclude that the classification of MS~1603.6+2600 depends strongly
on whether the X--ray flare observed by \scite{2001ApJ...561..938M}
was a type I X--ray burst or not.  If it was MS~1603.6+2600 is most
likely an ADC source. If the flare was just an X--ray flare and not a
type I X--ray burst MS~1603.6+2600 is either a nearby soft X--ray
transient in quiescence or a high inclination LMXB (possibly ADC)
source located well above the Galactic plane.

\section*{Acknowledgments} 
\noindent 
We would like to thank the referee for his/her useful comments and
suggestion which helped to improve the paper.  This research has made
use of the USNOFS Image and Catalogue Archive operated by the United
States Naval Observatory, Flagstaff Station
(http://www.nofs.navy.mil/data/fchpix/). PGJ is supported by EC Marie
Curie Fellowship HPMF--CT--2001--01308. MK is supported in part by a
Netherlands Organization for Scientific Research (NWO) grant.

\bibliographystyle{mn}

\end{document}